\documentclass[letter,11pt]{article}

\usepackage{float}
\usepackage{graphics,graphicx}
\usepackage{hyperref,setspace}

\begin{document}

\begin{center}
\includegraphics[width=\textwidth]{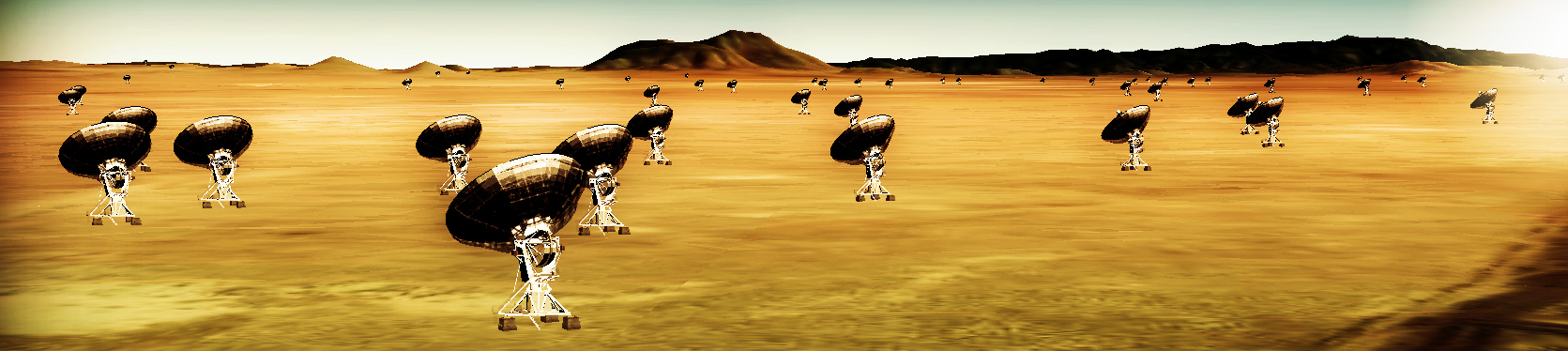}
\end{center}

\begin{center}

{\bf \Large Next Generation Very Large Array Memo No. 5}

\vspace{0.1in}

{\bf \Large Science Working Groups}

\vspace{0.1in}

{\bf \Large Project Overview}

\end{center}

\hrule 

\vspace{0.7cm}

\noindent C.L. Carilli$^{1,13}$,
M. McKinnon$^1$, J. Ott$^1$, A. Beasley$^2$,
A. Isella$^3$, E. Murphy$^4$, A. Leroy$^5$, C. Casey$^6$,
A. Moullet$^2$, M. Lacy$^2$, J. Hodge$^7$, G. Bower$^{8}$,
P. Demorest$^1$, C. Hull$^{9}$, M. Hughes$^{10}$, J. di Francesco$^{11}$,
D. Narayanan$^{12}$, B. Kent$^2$, B. Clark$^1$, B. Butler$^1$\\

\begin{center}
{\bf \large Abstract}\\
\end{center}

We summarize the design, capabilities, and some of the priority
science goals of a next generation Very Large Array (ngVLA). The ngVLA
is an interferometric array with 10 times larger effective collecting
area and 10 times higher spatial resolution than the current VLA and
the Atacama Large Millimeter Array (ALMA), optimized for operation in
the wavelength range 0.3cm to 3cm.  The ngVLA opens a new window on
the Universe through ultra-sensitive imaging of thermal line and
continuum emission down to milliarcecond resolution, as well as
unprecedented broad band continuum polarimetric imaging of non-thermal
processes.  The continuum resolution will reach 9mas at 1cm, with a
brightness temperature sensitivity of 6K in 1 hour. For spectral
lines, the array at $1''$ resolution will reach 0.3K surface
brightness sensitivity at 1cm and 10 km s$^{-1}$ spectral resolution
in 1 hour. These capabilities are the {\sl only} means with which to
answer a broad range of critical scientific questions in modern
astronomy, including direct imaging of planet formation in the
terrestrial-zone, studies of dust-obscured star formation and the
cosmic baryon cycle down to pc-scales out to the Virgo cluster, making
a cosmic census of the molecular gas which fuels star formation back
to first light and cosmic reionization, and novel techniques for
exploring temporal phenomena from milliseconds to years.  The ngVLA is
optimized for observations at wavelengths between the superb
performance of ALMA at submm wavelengths, and the future SKA-1 at few
centimeter and longer wavelengths.  This memo introduces the
project. The science capabilities are outlined in a parallel series of
white papers. We emphasize that this initial set of science goals are
simply a starting point for the project.  We invite comment on these
programs, as well as new ideas, through our public forum link on the
ngVLA web page:

\begin{center}
{\url{https://science.nrao.edu/futures/ngvla}}
\end{center}

\vspace{2mm} 

\begin{spacing}{1}
\footnotesize
\noindent $^1$NRAO, PO Box O, Socorro, NM \\
\noindent $^2$NRAO, 520 Edgemont Road, Charlottesville, VA \\
\noindent $^3$Rice University, Houston, TX \\
\noindent $^4$IPAC, Caltech, Pasadena, CA \\
\noindent $^5$Ohio State University, Columbus, OH \\
\noindent $^6$Univ. of Texas, Austin, TX \\
\noindent $^7$Leiden University, Leiden, Netherlands \\
\noindent $^{8}$ASIAA, Taipei, Taiwan, ROC \\
\noindent $^{9}$NRAO Jansky Fellow, Harvard-Smithsonian Center for Astrophysics, MA \\
\noindent $^{10}$Weslyan University, Middletown, CT \\
\noindent $^{11}$HIA, Victoria, BC, Canada \\
\noindent $^{12}$Haverford College, Haverford, PA \\
\noindent $^{13}$Cavendish Astrophysics Group, Cambridge, UK \\
\end{spacing}

\newpage

\tableofcontents

\newpage

\section{Introduction}

Inspired by the ground-breaking results coming from the Atacama Large
(sub)Millimeter Array, and the Jansky Very Large Array, the
astronomical community is considering a future large area radio array
optimized to perform imaging of thermal emission down to
milliarcsecond scales. Currently designated the `Next Generation Very
Large Array,' such an array would entail ten times the effective
collecting area of the JVLA and ALMA, operating from 1GHz to 115GHz,
with ten times longer baselines (300km) providing mas-resolution, plus
a dense core on km-scales for high surface brightness imaging. Such an
array bridges the gap between ALMA, a superb submillimeter array, and
the future Square Kilometer Array phase 1 (SKA-1), optimized for few
centimeter and longer wavelengths. The ngVLA opens unique new
parameter space in the imaging of thermal emission from cosmic objects
ranging from protoplanetary disks to distant galaxies, as well as
unprecedented broad band continuum polarimetric imaging of non-thermal
processes. 

We are considering the current VLA site as a possible location, in the
high desert plains of the Southwest USA. At over 2000m elevation, this
region provides good observing conditions for the frequencies under
consideration, including reasonable phase stability and opacity at 3mm
over a substantial fraction of the year (see JVLA and ngVLA memos by
Owen 2015, Clark 2015, Carilli 2015, Butler 2002).

Over the last year, the astronomical community has been considering
potential science programs that would drive the design of a future
large area facility operating in this wavelength range.  These goals
are described in a series of reports published as part of the ngVLA
memo series, and can be found in the ngVLA memo series:

\begin{center}
{\url{http://library.nrao.edu/ngvla.shtml}}
\end{center}

\begin{itemize}

\item Isella et al., 2015, 'Cradle of Life' (ngVLA Memo 6)

\item Leroy et al., 2015, 'Galaxy Ecosystems' (ngVLA Memo 7)

\item Casey et al. 2015, 'Galaxy Assembly through Cosmic Time' (ngVLA Memo 8)

\item Bower et al. 2015, 'Time Domain, Cosmology, Physics' (ngVLA Memo 9)

\end{itemize}

The white papers will be expanded with new ideas, and more detailed
analyses, as the project progresses (eg. a white paper on
magneto-plasma processes on scales from the Sun to clusters of
galaxies is currently in preparation).  In the coming months, the
project will initiate mechanisms to further expand the ngVLA science
program, through continued community leadership.

Such a facility will have broad impact on many of the paramount
questions in modern astronomy. The science working groups are in the
process of identifying a number of key science programs that push the
requirements of the telscope. Three exciting programs that have
come to the fore thus far, and {\sl that can only be done with the
ngVLA}, include:

\begin{itemize}

\item {\bf Imaging the 'terrestrial-zone' of planet formation in
protoplanetary disks}: Probing dust gaps on 1AU scales at the distance
of the nearest major star forming regions (Taurus and Ophiucus
distance $\sim$ 130pc) requires baselines 10 times that of the JVLA,
with a sensitivity adequate to reach a few K brightness at 1cm
wavelength and 9mas resolution. Note that these inner regions of
protoplanetry disks are optically thick at shorter wavelengths (see
section 4.1).  The ngVLA will image the gap-structures indicating
planet formation on solar-system scales, determine the growth of
grains from dust to pebbles to planets, and image accretion onto the
proto-planets themselves.

\item {\bf ISM and star formation physics on scales from GMCs down to
cloud cores throughout the local super-cluster}: a centrally condensed
antenna distribution on scales of a few km (perhaps up to 50\%
of the total collecting area), is required for wide field, high
surface brightness (mK) sensitivity.  The ngVLA covers the spectral
range richest in the ground state transitions of the most important
molecules in astrochemistry and astrobiology, as well as key thermal
and non-thermal continuum emission process relating to star
formation. The ngVLA will perform wide field imaging of line and
continuum emission on scales from GMCs (100pc) down to clump/cores
(few pc) in galaxies out to the Virgo Cluster.

\item {\bf A complete census of the cold molecular gas fueling the
star formation history of the Universe back to the first galaxies:}
octave bandwidth at $\sim 1$cm wavelength, is required for large
cosmic volume surveys of low order CO emission from distant galaxies
(the fundamental tracer of total gas mass), as well as for dense gas
tracers such as HCN and HCO+.  The spatial resolution and sensitivity
will also be adequate to image gas dynamics on sub-kpc scales and detect
molecular gas masses down to dwarf galaxies.

\end{itemize}

In this summary paper, we present a general description of the
project, basic design goals for sensitivity and resolution, and the
unique observational parameter space opened by such a revolutionary
facility.  We emphasize that the ngVLA is a project under
development. While the broad parameter space is reasonably well
delineated, there are many issues to explore, ranging from element
diameter to the number of frequency bands to the detailed array
configuration, including consideration of VLBI-length baselines (see
section 2.2). The science white papers are identifying the primary
science use cases that will dictate the ultimate design of the
telescope, in concert with the goal of minimization of construction
and operations costs. The requirements will mature with time, informed
by ALMA, the JVLA, the imminent JWST and thirty meter-class optical
telescopes, and others.

\section{Telescope specifications}

\subsection{Basic array}

In Table 1 we summarize the initial telescope specifications for the
ngVLA. As a first pass, we present numbers for an 18m diameter
antenna, although the range from 12m to 25m is being considered.  A
key design goal is good antenna performance at higher frequency,
eg. at least 75\% efficiency at 30GHz.  The nominal frequency range of
1GHz to 115GHz is also under discussion. The bandwidths quoted are
predominantly 2:1, or less, although broader bandwidths are being
investigated. Receiver temperatures are based on ALMA and VLA
experience. We emphasize that these specifications are a first pass at
defining the facility, and that this should be considered an evolving
study.

Brightness sensitivity for an array is critically dependent on the
array configuration.  We are assuming an array of 300 antennas in this
current configuration.  The ngVLA has the competing desires of both
good point source sensitivity at full resolution for few hundred km
baselines, and good surface brightness sensitivity on scales
approaching the primary beam size. Clark \& Brisken (2015) explore
different array configurations that might provide a reasonable
compromise through judicious weighting of the visibilities for a given
application (see eg. Lal et al. 2010 for similar studies for the
SKA). It is important to recognize the fact that for any given
observation, from full resolution imaging of small fields, to imaging
structure on scales approaching that of the primary beam, some
compromise will have to be accepted. 

For the numbers in Table 1, we have used the Clark/Conway
configurations described in ngVLA memos 2 and 3. Very briefly, this
array entails a series of concentric 'fat-ring' configurations out to
a maximum baseline of 300km, plus about 20\% of the area in a compact
core in the inner 300m.  The configuration will be a primary area for
investigation in the coming years. We have investigated different
Briggs weighting schemes for specific science applications, and find
that the Clark/Conway configuration provides a reasonable starting
compromise for further calculation (see notes to Table 1).

\begin{table}
\footnotesize
\caption{Next Generation VLA nominal parameters}
\label{tlab}
\begin{tabular}{lccccc}\hline
~ & 2GHz & 10GHz & 30GHz & 80GHz & 100GHz \\ 
\hline
Field of View FWHM (18m$^a$) arcmin & 29 & 5.9 & 2 & 0.6 & 0.51 \\
Aperture Efficiency (\%) &  65 & 80 & 75 & 40 & 30  \\
A$_{eff}^b$  x$10^4$  m$^2$ & 5.1 & 6.2 & 5.9 & 3.1 & 2.3    \\
T$_{sys}^c$ K & 29 & 34 & 45 & 70 & 80 \\
Bandwidth$^d$ GHz & 2 & 8 & 20 & 30 & 30 \\
Continuum rms$^e$ 1hr, $\mu$Jy bm$^{-1}$ & 0.93 & 0.45 & 0.39 & 0.96 & 1.48 \\
Line rms 1hr, 10 km s$^{-1}$, $\mu$Jy bm$^{-1}$ & 221 & 70 & 57 & 100 & 130 \\
Resolution$^f$  FWHM milliarcsec & 140 & 28 & 9.2 & 3.5 & 2.8 \\
T$_B^g$ rms continuum 1hr  K & 14 & 7 & 6 & 15 & 23 \\
Line$^h$ rms 1hr, $1"$, 10 km s$^{-1}$, $\mu$Jy bm$^{-1}$ & 340 & 140 & 240 & 860 & -- \\
T$_B^i$ rms line, 1hr, $1"$, 10 km s$^{-1}$,  K & 100 & 1.8 & 0.32 & 0.17 & --  \\
\hline
\vspace{0.1cm}
\end{tabular}
$^a$Under investigation: antenna diameters from 12m to 25m are being considered. \\ 
$^b$300 x 18m antennas with given efficiency. \\
$^c$Current performance of JVLA below 50GHz. Above 70GHz we assume the T$_{sys}$ =60K value
for ALMA at 86GHz, increased by 15\% and 25\%, respectively, due to 
increased sky contribution at 2200m. \\
$^d$Under investigation. For much wider bandwidths, system temperatures are 
likely to be larger. \\
$^e$Noise in 1hour for given continuum bandwidth for a Clark/Conway configuration  
(ngVLA memo 2 and 3) scaled to a maximum baseline of 300km,
using Briggs weighting with R=0. Using R=1 decreases the noise by a factor 0.87, 
and using R=-1 increases the noise by a factor 2.5. \\
$^f$Synthesized beam for a Clark/Conway configuration scaled to a
maximum baseline of 300km, using Briggs weighting with R=0.  For R=1, the beam size increases
by a factor 1.36, and for R=-1 the beam size decreases by a factor 0.63. \\
$^g$Continuum brightness temperature corresponding to point source sensitivity (row 6) and resolution of Clark/Conway configuration, using Briggs weighting with R = 0 (row 8).  \\
$^h$Line rms in 1hr, 10 km s$^{-1}$, after tapering to $1"$ resolution for the Clark/Conway configuration. \\
$^i$Line brightness temperature rms in 1hr, 10 km s$^{-1}$, after tapering to $1"$ resolution for the Clark/Conway configuration. \\
\end{table}

\subsection{VLBI implementation}

The science white papers present a number of compelling VLBI
astrometric science programs made possible by the increased
sensitivity of the ngVLA.  These include: Local Group cosmology
through measurements of proper motions of nearby galaxies, delineation
of the full spiral structure of the Milky Way, and measuring the
masses of supermassive black holes and H$_0$.

The exact implementation of interferometry with the
ngVLA on baselines longer than the nominal 300km array remains under
investigation.  These astrometric programs require excellent
sensitivity per baseline, but may not require dense coverage of the UV
plane, since high dynamic range imaging may not be required.

One possible implementation would be to use the ngVLA as an
ultra-sensitive, anchoring instrument, in concert with radio 
telescopes across the globe. Such a model would parallel the
planned implementation for submm VLBI, which employs the
ultra-sensitive phased ALMA, plus single dish submm telescopes around
the globe, to perform high priority science programs, such as imaging
the event horizons of supermassive black holes (Akiyama et
al. 2015). A second possibility would be to include out-lying stations
within the ngVLA construction plan itself, perhaps comprising up to
20\% of the total area out to trans-continental baselines. The cost,
practicability, and performance of different options for VLBI will be
studied in the coming year.

\section{New Parameter Space}

Figure 1 shows one slice through the parameter space covered by the
ngVLA: resolution versus frequency, along with other existing and
planned facilities.  The maximum baselines of the ngVLA imply a
resolution of better than 10mas at 1cm. As we shall see below, coupled
with the high sensitivity of the array, this resolution provides a
unique window into the formation of planets in disks on scales of our
own Solar system at the distance of the nearest active star forming
regions, eg. Taurus and Ophiucus.

Figure 2 shows a second slice through parameter space: effective
collecting area versus frequency.  In this case, we have not included
much higher and lower frequencies, eg. the SKA-1 will extended
to much lower frequency (below 100MHz, including SKA-Low), while ALMA
extends up to almost a THz.

\begin{figure}[tbh]
\begin{center}
\includegraphics[scale=0.32]{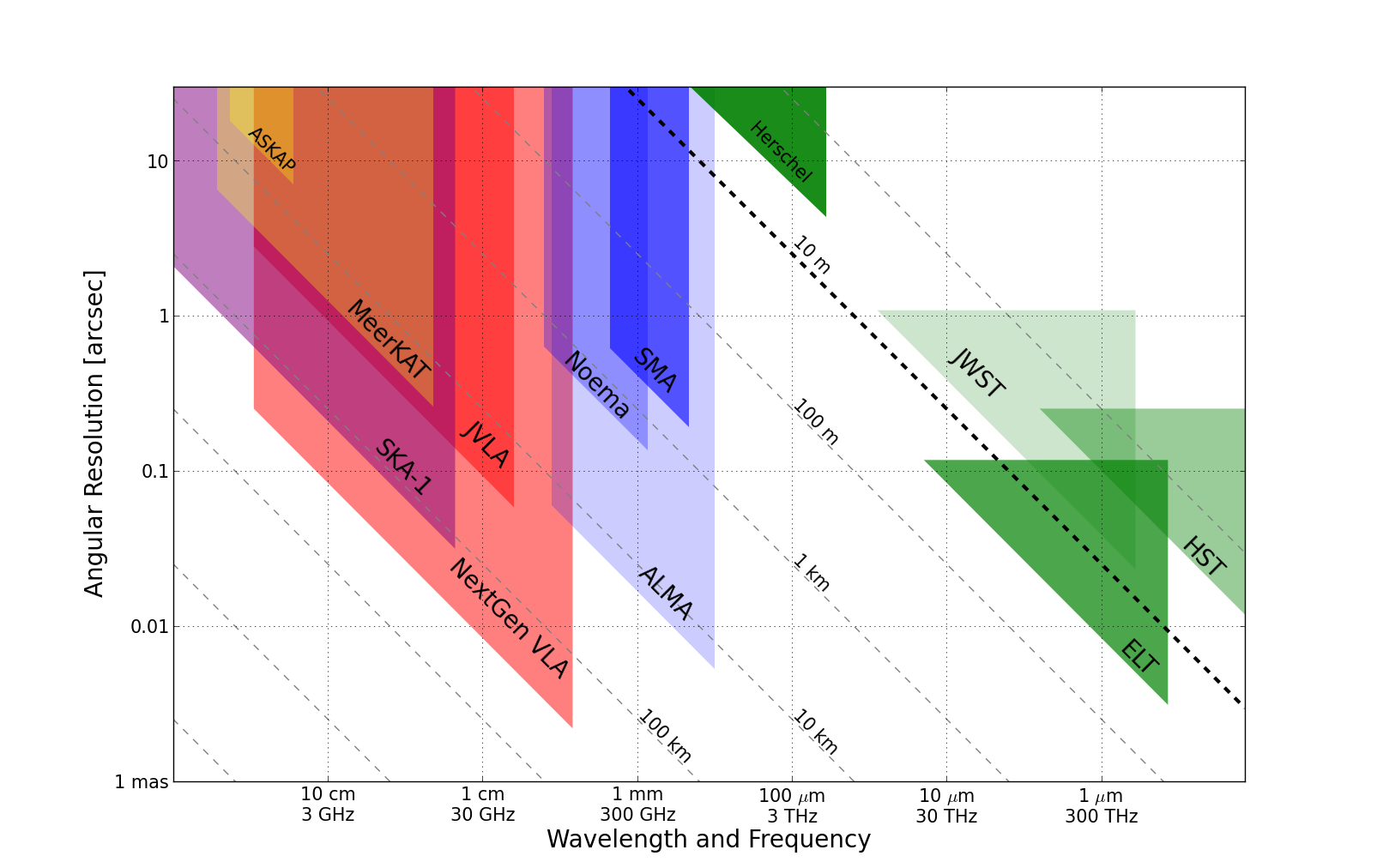}
\end{center}
\caption{\footnotesize \em{Spatial resolution versus frequency set by the 
maximum baselines of the ngVLA, and
other existing and planned facilities across a broad range of
wavelengths.  }}
\end{figure}

\begin{figure}[tbh]
\begin{center}
\includegraphics[scale=0.32]{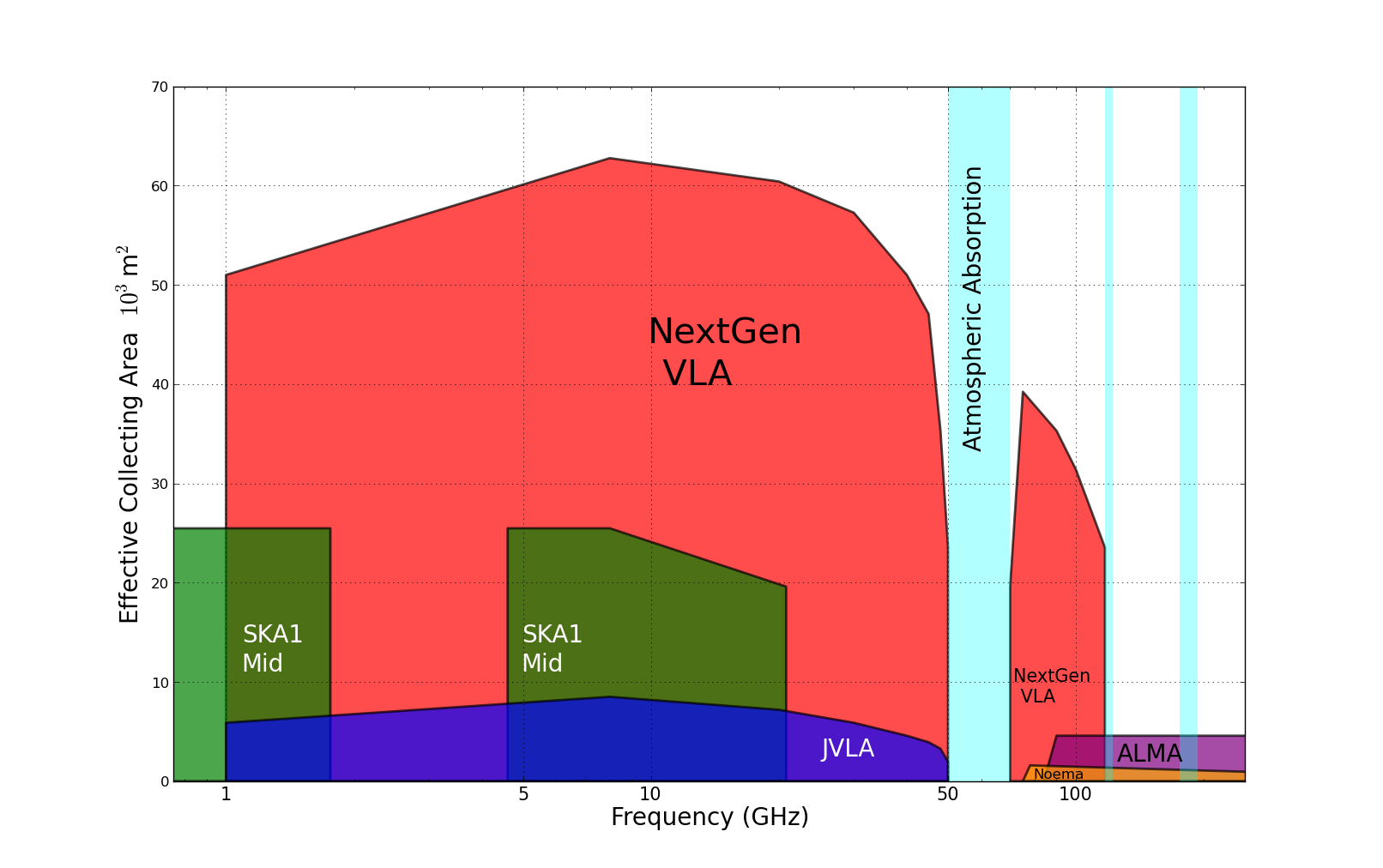}
\end{center}
\caption{\footnotesize \em{Effective collecting area versus frequency for the ngVLA,
and other existing or planned facilities operating in a comparable
frequency range. We have not included much higher and lower
frequencies, eg. the SKA-1 will extended to below 100MHz
(including SKA-Low), while ALMA extends up to close to a THz. }}
\end{figure}

Given the collecting area and reasonable receiver performance (Table
1), the ngVLA will achieve sub-$\mu$Jy sensitivity in the continuum in
1 hour at 1cm (30GHz). This implies that, at 1cm, the ngVLA will
obtain 6K brightness temperature sensitivity with 9mas resolution in
just 1 hour!

We note that there are other aspects of telescope phase space that are
relevant, including field of view and mapping speed, configuration and
surface brightness sensitivity, bandwidth, T$_{sys}$, etc...  Given
the early stage in the design, we have presented the two principle and
simplest design goals, namely, maximum spatial resolution and total
effective collecting area. A deeper consideration of parameter space
will depend on the primary science drivers that emerge in the coming
years.

\section{Science Examples}

In the following, we highlight some of the science that is enabled by
such a revolutionary facility. These three areas are among the high
priority goals identified by the science working groups, and in
particular, these are the goals that have been best quantified to
date.  We note that the most important science from such a
revolutionary facility is difficult to predict, and perhaps the most
important aspect of the science analysis is simply the large volume of
unique parameter space opened by the ngVLA (Figs 1 and 2).

\subsection{Imaging terrestrial-zone planet formation}

With the discovery of thousands of extrasolar planets, and the first
high resolution images of protoplanetary disks with ALMA, the field of
extrasolar planets and planet formation has gone from rudimentary
studies, to a dominant field in astrophysics, in less than a
decade. This remarkable progress promises to continue, as ALMA comes
into full operation, and with future space missions targetting planet
detection, such as the High Definition Space Telescope, for which the
primary science goals are direct imaging of terrestrial planets and
the search for atmospheric bio-signatures.

The first high resolution images from ALMA of the protoplanetary disk
in HL Tau are clearly game-changing (Brogan et al. 2015).  The ALMA
images show a dust disk out to 100AU radius, with a series of gaps at
radii ranging from 13 AU to 80AU. These gaps may correspond to
the formation zones of planets. Coupled with JVLA imaging at longer
wavelengths, these HL Tau images usher in a new era in the study of
planet formation.

While revolutionary, there are limitations to the current
capabilities of ALMA and the JVLA in the study of protoplanetary
disks. First, for ALMA, the inner 10AU of protoplanetary disks like HL
Tau become optically thick at wavelengths of 3mm and shorter. Second,
for the JVLA, the sensitivity and spatial resolution are insufficient
to image the terrestrial-zone of planet formation at the longer
wavelengths where the disks become optically thin.

\begin{figure}[tbh]
\begin{center}
\includegraphics[width=\textwidth]{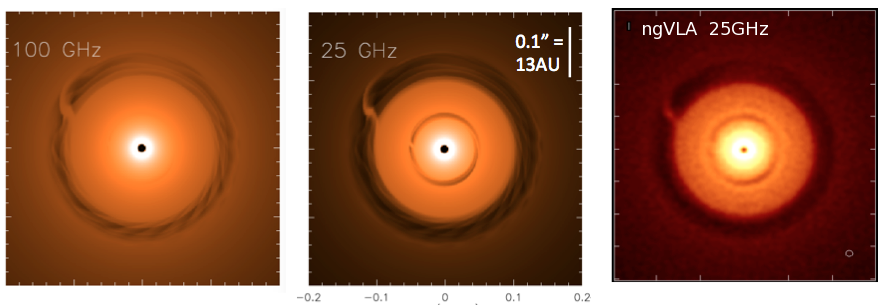}
\end{center}
\caption{\footnotesize \em{Models and images of a $\sim 1$Myr old protoplanetary
disk, comparable to HL Tau, at a distance of 130pc. This 'minimum mass
solar nebula disk' has a mass of 0.1M$_\odot$ orbiting a 1 M$_\odot$ star.
The model includes the formation of a Jupiter
mass planet at 13AU radius, and Saturn at 6AU.
The left frame shows the model emission at 100GHz, the center frame
shows the 25GHz model, and the right shows the ngVLA image for a
100hour observation at 25GHz with 10mas resolution. The noise 
in the ngVLA image is 0.1$\mu$Jy, corresponding to 1K at 10mas resolution. }}
\end{figure}

The ngVLA solves both of these problems, through ultra-high sensitivity
in the 0.3cm to 3cm range, with milliarcsecond resolution. Figure 3
shows a simulation of the ability of the ngVLA to probe the previously
inaccessible scales of 1AU to 10AU. This simulation involves an
HL-Tau like protoplanetary disk, including the formation of a Jupiter
mass planet at 13AU radius, and Saturn at 6AU. Note that the inner
ring caused by Saturn is optically thick at 3mm.  However, this inner
gap is easily visible at 25GHz, and well imaged by the
ngVLA. Moreover, the ngVLA will have the sensitivity and resolution to
image circum-planetary disks, ie. the formation of planets themselves
via accretion. In parallel, the ngVLA covers the optimum frequency
range to study pre-biotic molecules, including rudimentary amino acids
such as glycine (see Isella et al. 2015 for more details).

{\sl Next Generation Synergy:} The High Definition Space
Telescope has made its highest priority goals the direct imaging of
terrestrial-zone planets, and detection of atmospheric biosignatures.
The ngVLA provides a perfect evolutionary compliment to the HDST
goals, through unparalleled imaging of terrestrial zone planet
formation, and the study of pre-biotic molecules.

\subsection{The dense gas history of the Universe}

Using deep fields at optical through radio wavelengths, the evolution
of cosmic star formation and the build up of stellar mass have been
determined in exquisite detail, from the epoch of first light (cosmic
reionization, $z > 7$), through the peak epoch of cosmic star
formation ('epoch of galaxy assembly', $z \sim 1$ to 3), to the
present day (Madau \& Dickinson 2014). However, these studies reveal
only one aspect of the baryonic evolution of galaxies, namely, the
stars. What is currently less well understood, but equally important,
is the cosmic evolution of the cool, molecular gas out of which stars
form.  Initial in-roads into the study of the cool gas content of
galaxies has been made using the JVLA, GBT, Plateau de Bure, and now
ALMA. These initial studies have shown a profound change in the
baryonic content of star forming galaxies out to the epoch of galaxy
assembly: the gas baryon fraction (the gas to stellar mass ratio)
increases from less than 10\% nearby, to unity, or larger, at $z \sim
2$ to 3 (Genzel et al. 2015, Carilli \& Walter 2013).  
This profound change in galaxy properties with redshift is
likely the root-cause of the evolution of the cosmic star formation
rate.

\begin{figure}[tbh]
\begin{center}
\includegraphics[width=\textwidth]{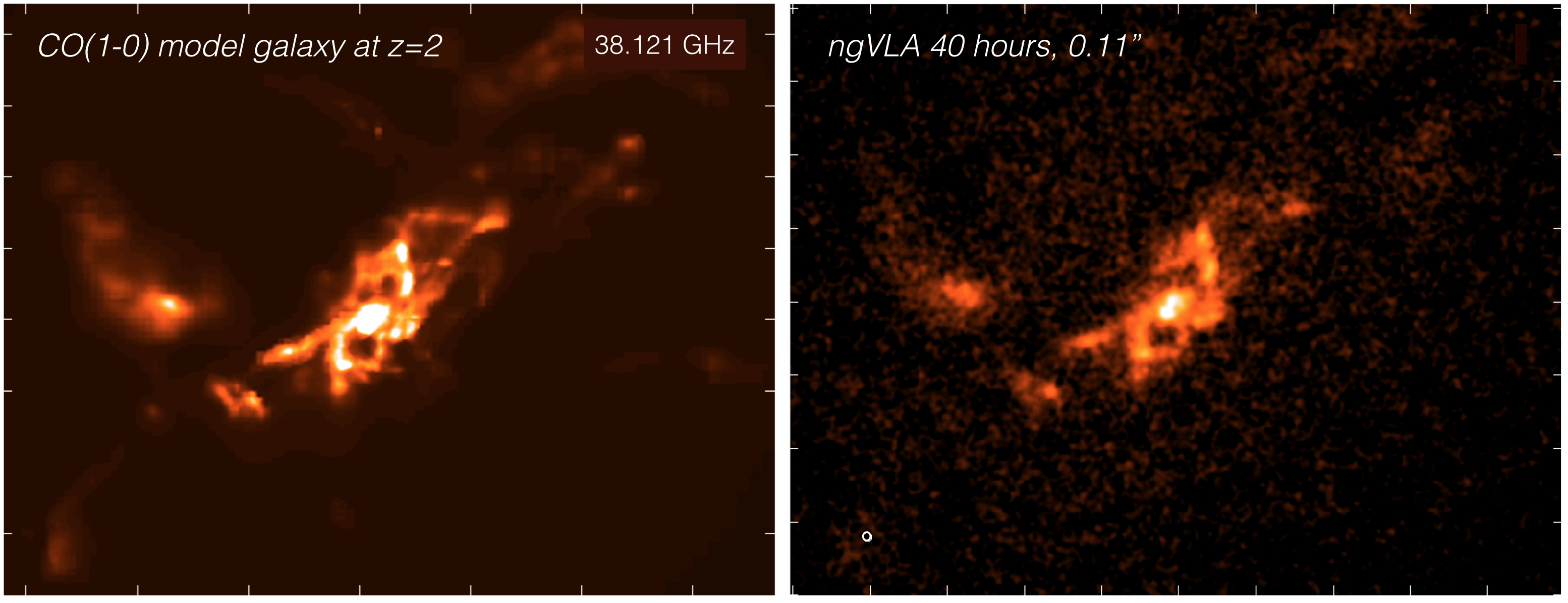}
\end{center}
\caption{\footnotesize \em{
Left: A model of the integrated CO 1-0 emission from a
massive z=2 galaxy from the cosmological zoom simulations of Narayanan
et al. (2015). The total SFR $= 150$ M$_\odot$ year$^{-1}$, and the stellar
mass = $4\times 10^{11}$ M$_\odot$.  The native resolution (pixel size)
is 30mas, and the peak brightness temperature is 14K. The fainter
regions have T$_B \ge 0.1$K. Right: the ngVLA image of
the field assuming a 8 x 5hour synthesis using only antennas within a
15km radius (about 50\% of the full array for the Clark/Conway
configuration, and using Briggs weighting
with R=0.5. The rms noise is 5$\mu$Jy beam$^{-1}$, and the beam size
is $0.11"$. One tick mark = $1"$. The peak surface brightness is 0.18
mJy beam$^{-1}$. 
}}
\end{figure}

However, studies of the gas mass in early galaxies, typically using
the low order transitions of CO, remain severely sensitivity limited,
requiring long observations even for the more massive galaxies.
The sensitivity and resolution of the ngVLA opens a new window on the
gas properties of early galaxies, through efficient large cosmic
volume surveys for low order CO emission, and detailed imaging of gas
in galaxies to sub-kpc scales (see Casey et al. 2015).  The ngVLA will
detect CO emission from tens to hundreds of galaxies per hour in
surveys in the 20GHz to 40GHz range. In parallel, imaging of the gas
dynamics will allow for an empirical calibration of the CO luminosity
to gas mass conversion factor at high redshift.

Figure 4 shows a simulation of the CO 1-0 emission from a 
massive z=2 galaxy from the cosmological zoom simulations of Narayanan
et al. (2015), plus the ngVLA simulated image. The ngVLA reaches an
rms noise of 5$\mu$Jy beam$^{-1}$ (over 9MHz bandwidth and 40hours),
and the beam size is $0.11"$ = 0.9kpc at z=2, only using antennas
within 15km radius of the array center. The ngVLA can detect the large
scale gas distribution, including tidal structures, streamers,
satellite galaxies, and possible accretion. Note that the rms
sensitivity of the ngVLA image corresponds to an H$_2$ mass
limit of $3.3\times 10^8$ ($\alpha$/4) M$_\odot$.  Further, the ngVLA
has the resolution to image the gas dynamics on scales approaching
GMCs. For comparison, the JVLA in a similar integration time would
only detect the brightest two knots at the very center of galaxy,
while emission from the high order transitions imaged by ALMA misses
the extended, low excitation, diffuse gas in the system. 

{\sl Next Generation Synergy:} With new facilities such as
thirty-meter class optical telescopes, the JWST, and ALMA, study of
the stars, ionized gas, and dust during the peak epochs of galaxy
formation, will continue to accelerate. The ngVLA sensitivity and
resolution in the 0.3cm to 3cm window is the required complement to
such studies, through observation of the cool gas out of which stars
form throughout the Cosmos.

\subsection{Ultra-sensitive, wide field imaging}

Science working group 2 (`Galaxy ecosystems'; Leroy etal. 2015)
emphasized the extraordinary mapping speed of the ngVLA in line and
continuum, for study of the gas and star formation in the nearby
Universe.  The frequency range of the ngVLA covers, simultaneously,
multiple continuum emission mechanisms, from synchrotron, to
free-free, to cold (or spinning) dust. These mechanisms are key
diagnostics of star formation, cosmic rays, magnetic fields, and other
important ISM properties. This range also covers low order and maser
transitions of most astrochemically important molecules, such as CO,
HCN, HCO$^+$, NH$_3$, H$_2$O, CS...
 
Figure 5 shows an ngVLA simulation of the thermal free-free emission
in the 30GHz band from a star forming galaxy at 27Mpc distance, with a
moderate star formation rate of 4 M$_\odot$ year$^{-1}$.  The ngVLA
will image the free-free emission with a sensitivity adequate to
detect an HII region associated with a single O7.5 main sequence star
at the distance of the Virgo cluster!  In general, the combination of
spectral and spatial resolution will allow for decomposition
of the myriad spectral lines, and various continuum emission
mechanisms, on scales down to a few parsecs at the distance of Virgo,
thereby enabling Local-Group-type science throughout the local
supercluster.

\begin{figure}[tbh]
\begin{center}
\includegraphics[width=\textwidth]{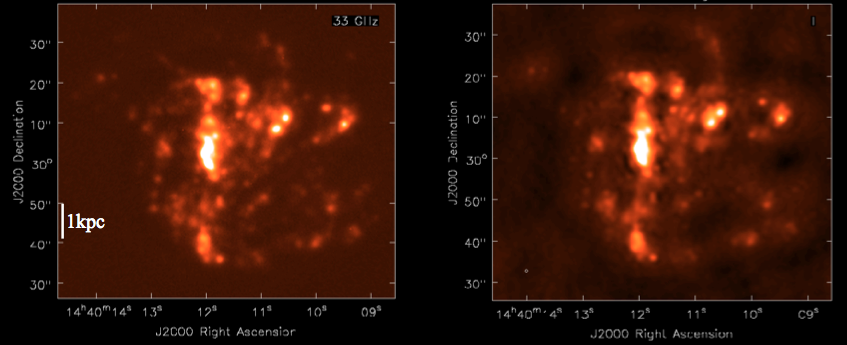}
\end{center}
\caption{\footnotesize \em{Left: a model for the thermal free-free
emission from NGC 5713 at a distance of 27Mpc with 
a SFR = 4 M$_\odot$ year$^{-1}$. The model was estimated from H$\alpha$ imaging
at a native resolution of 2$"$. The peak brightness temperature is 
150mK, and the fainter knots are about 1mK. Right: The ngVLA image
for 10hrs integration, with a bandwidth of 20GHz, centered at 30GHz. 
The rms is 0.1$\mu$Jy beam$^{-1}$. 
Note that the ngVLA image has been restored with a beam of 
$0.5"$. }}
\end{figure}

\subsection{Exploring the Time Domain}

The ngVLA is being designed for optimal exploitation of the time
domain. Fast triggered response modes on minute timescales will be
standard practice. Commensal searches for ultra-fast transients, such
as Fast Radio Bursts or SETI signals, will also be incorporated into
the design. And monitoring of slow transients, from novae to AGN, will
be possible at unprecedented sensitivities, bandwidths, and angular
resolutions. The 2cm and shorter capabilities will be complimentary to
the SKA-1 at longer wavelengths, in particular for the broad band
phenomena typical of fast and slow transients.

The broad band coverage and extreme sensitivity of the ngVLA provides
a powerful tool to search for, and characterize, the early time
emission from processes ranging from gravity wave EM counter-parts to
tidal disruption events around supermassive black holes as well as
probing through the dense interstellar fog in search of Galactic
Center pulsars. The system will also provide unique insights into
variable radio emission associated with 'exo-space weather,' such as
stellar winds, flares, and aurorae.  Moreover, many transient phenomena
peak earlier, and brighter, at higher frequencies, and full spectral
coverage to high frequency is required for accurate calorimetry. Full
polarization information will also be available, as a key diagnostic
on the physical emission mechanism and propagation effects. 

\vspace{0.5cm}

We invite the reader to investigate the science programs in more
detail in the working group reports, as well as to participate in the
public forums and meetings in the on-going development of the ngVLA
science case.

\section*{Acknowledgments}

The National Radio Astronomy Observatory is a facility of the National
Science Foundation operated under cooperative agreement by Associated
Universities, Inc.

\vskip 0.2in

\noindent{\sl References}

\noindent Akiyama, K. et al. 2015, ApJ, 807, 150

\noindent Bower, G. et al.  2015, {\sl Next Generation VLA memo. No. 9}

\noindent Brogan, C. et al. 2015, ApJ, 808, L3

\noindent Butler, B. 2002, VLA Test Memo 232

\noindent Carilli, C. 2015, {\sl Next Generation VLA memo. No. 1}

\noindent Carilli, C. \& Walter, F. 2013, ARAA, 51, 105

\noindent Casey, C. et al.  2015, {\sl Next Generation VLA memo. No. 8}

\noindent Clark, B. \& Brisken, W. 2015, {\sl Next Generation VLA memo. No. 3}

\noindent Clark, B. 2015, {\sl Next Generation VLA memo. No. 2}

\noindent Genzel, R. et al. 2015, ApJ, 800, 20

\noindent Isella, A. et al.  2015, {\sl Next Generation VLA memo. No. 6}

\noindent Lal, D., Lobanov, A., Jimenez-Monferrer, S. 2011, 
SKA Design Studies Technical Memo 107

\noindent Madau, P. \& Dickinson, M. 2014, ARAA, 52, 415

\noindent Leroy, E. et al.  2015, {\sl Next Generation VLA memo. No. 7}

\noindent Narayanan, D. et al. 2015, Nature, 525, 496

\noindent Owen, F. 2015, {\sl Next Generation VLA memo. No. 4}

\end{document}